# Thermodynamic modeling of the LiCl-KCl-$LaCl_3$ system with Bayesian model selection and uncertainty quantification

Rushi Gong[1,*], Shun-Li Shang[1,*], Vitaliy G. Goncharov[2,3], Cillian Cockrell[4,5], Kostya Trachenko[5], Paul A. Bingham[6], Xiaofeng Guo[2,3], and Zi-Kui Liu[1]

[1] *Department of Materials Science and Engineering, The Pennsylvania State University, University Park, PA 16802, United States*

[2] *Department of Chemistry, Washington State University, Pullman, Washington, 99164, United States*

[3] *Materials Science and Engineering Program, Washington State University, Pullman, Washington 99164, United States*

[4] *Nuclear Futures Institute, Bangor University, Bangor, LL57 1UT, United Kingdom*

[5] *School of Physical and Chemical Sciences, Queen Mary University of London, Mile End Road, London, E1 4NS, United Kingdom*

[6] *Materials and Engineering Research Institute, Sheffield Hallam University, Sheffield, S1 1WB, United Kingdom*

*Corresponding authors

Corresponding emails: rfg5281@psu.edu (R. S. Gong), sus26@psu.edu (S. L. Shang)

**Abstract**

Chloride molten salts are increasingly used in pyroprocessing techniques for the separation of lanthanides. Understanding thermodynamic properties of these salts is essential to predict their critical characteristics and optimize the separation process. Several thermodynamic models, including the associate model, the two-sublattice ionic model, and the modified quasichemical model with quadruplet approximation (MQMQA), have been utilized in the literature to capture the complexity of molten salts. In the present work, the Bayes factor is used to guide the model selection process for thermodynamic modeling of the $KCl-LaCl_3$ system and provide statistical comparison of various models. The results indicate that the MQMQA model is the most favorable one based on available data. The $LiCl-KCl-LaCl_3$ system has been further modelled with uncertainty quantification (UQ) using MQMQA with the thermodynamic properties of compounds in $KCl-LaCl_3$ predicted by the quasiharmonic approach in terms of first-principles phonon calculations as a function of temperature. The calculated phase stability shows excellent agreement with experimental data, indicating that an appropriate thermodynamic model is important for accurately predicting critical characteristics of complex molten salts.

**Highlights:**

- Temperature-dependent thermodynamic properties for compounds in KCl-$LaCl_3$ by DFT-based phonon calculations.
- Bayes factor to select thermodynamic model for the KCl-$LaCl_3$ liquid.
- Thermodynamic modeling with UQ of LiCl-KCl-$LaCl_3$ with MQMQA used for liquid.

## 1 Introduction

Chloride molten salts have gained significant attention for their potential applications in advanced nuclear reactors and pyroprocessing techniques [1–3]. Effective separation of fission products, such as lanthanides using pyroprocessing techniques, requires a comprehensive understanding of their thermodynamic properties of molten salts. The CALPHAD (CALculation of PHAse Diagram) modeling approach [4–6] is an effective method for investigating phase stability and thermodynamic properties in multicomponent systems. For modeling of complex molten salts, various thermodynamic models within the CALPHAD framework have been developed to describe intricate behaviors such as short-range ordering. These models include the associate model [7], the two-sublattice ionic model [8], and the modified quasichemical model with quadruplet approximation (MQMQA) [9,10]. All of them have been applied in CALPHAD modeling of molten salts [11–16]. However, systematical comparison of these models and selection of the most appropriate model for molten salts remain challenges, due to different physical interpretations and the complexities involved in quantifying model performance.

Several model selection criteria have been employed in the CALPHAD modeling approach. For instance, the corrected Akaike information criterion (AICc) [17] has been implemented in the open-source software ESPEI [18], facilitating the comparison and selection of Gibbs energy models. Shang et al. [19] utilized the AICc to select models aiming at fitting phonon and thermal electron contributions to free energy. Paulson et al. [20] used Bayes factor [21] to select heat capacity models for the alpha, beta and liquid phases of hafnium, while Honarmandi et al. [22]

used Bayes factor and Bayesian model average to compare models in the Hf-Si system. These studies highlight the utility of statistical comparison in guiding model selection in CALPHAD modeling.

In the present work, thermodynamic modeling of the LiCl-KCl-$LaCl_3$ system was performed using experimental data in the literature, supplemented with thermochemical data of compounds predicted by first-principles calculations based on density functional theory (DFT). The open-source software tools of ESPEI [18] and PyCalphad [23] were employed for the modeling process. The recent implementation of MQMQA [24] facilitates Bayesian parameter estimation, uncertainty quantification (UQ), and Bayesian model selection in modeling molten salts [16]. Four candidate thermodynamic models for the KCl-$LaCl_3$ liquid were compared in the present work. The optimal model for this system is identified using Bayes factor, leading to a further model of the ternary LiCl-KCl-$LaCl_3$ system, which demonstrates an excellent agreement with experimental data reported in the literature. The present work compares thermodynamic models commonly used in CALPHAD modeling for liquid, providing insights into an effective model selection strategy.

## 2 Literature review of the KCl-$LaCl_3$ and LiCl-KCl-$LaCl_3$ systems

The LiCl-KCl-$LaCl_3$ system contains the liquid phase, three binary compounds of LiCl, KCl, and $LaCl_3$, and two ternary compounds of $K_2LaCl_5$ and $K_3La_5Cl_{18}$ as summarized by Hao et al. [14]. $K_2LaCl_5$ is with a Pnma structure measured by Meyer et al. [25]. Seifert et al. [26] determined the structure of $K_3La_5Cl_{18}$ with a space group of $P6_3/m$ through X-ray diffraction. The values of formation enthalpy of $K_2LaCl_5$ and $K_3La_5Cl_{18}$ were measured by Seifert et al. [26] using solution calorimetry. Reuter and Seifert [27] reported the values of heat capacity of $K_2LaCl_5$ and $K_3La_5Cl_{18}$ using differential scanning calorimetry (DSC). Gaune-Escard and Rycerz [28] also measured the heat capacity of $K_3La_5Cl_{18}$ using DSC. Papatheodorou and Ostvold [29] reported mixing enthalpy in liquid in KCl-$LaCl_3$ through calorimetric experiments. Qiao et al. [30] utilized the differential thermal analysis (DTA) techniques to determine melting and phase transition temperatures. In the KCl-$LaCl_3$ system, Song and Zheng [31] measured liquidus by DTA. Seifert et al. [26] measured phase boundaries in the $LaCl_3$-rich range by DTA.

In the ternary LiCl-KCl-$LaCl_3$ system, Bagri and Simpson [32] and Samin et al. [33] reported activity coefficient values for $LaCl_3$ in molten LiCl-KCl eutectic salt using electromotive force measurements and cyclic voltammetry, respectively. Regarding the phase diagram, Song and Zheng [31] reported the liquidus projection and six isopleths. Two research works, by Nakamura el al. [34] and Krishnan et al. [35], constructed the pseudo-binary phase diagram from the LiCl-KCl eutectic to 25 mol% of $LaCl_3$ in the LiCl-KCl eutectic through DSC. Mixing enthalpies of $LaCl_3$ in LiCl-KCl eutectic with the $LaCl_3$ concentration from 0 to ~80 mol% have also been studied by Goncharov et al. [36], using both high-temperature drop calorimetry and *ab initio* molecular dynamics. Previously, high-temperature drop calorimetry was also used for examining other similar molten chloride systems [36–38], with recent revisions in the experimental method [39,40]. These newly obtained mixing enthalpy values of $LaCl_3$-(LiCl-KCl) were used in this work for benchmarking our modeling results.

# 3 Methodology

## 3.1 DFT-based first-principles calculations

### *3.1.1 Helmholtz energy at finite temperatures*

The Helmholtz energy $F(V,T)$ as a function of volume ($V$) and temperature ($T$) in terms of the DFT-based quasiharmonic approximation (QHA) can be determined by [41],

$$F(V,T) = E(V) + F_{el}(V,T) + F_{vib}(V,T) \qquad \textbf{\textit{Eq. 1}}$$

where $E(V)$ is static energy at 0 K without the zero-point vibrational energy, and $F_{el}(V,T)$ and $F_{vib}(V,T)$ represent the temperature-dependent thermal electron contribution and the vibrational contribution, respectively. In the present work, a four-parameter Birch-Murnaghan (BM4) equation of state (EOS) [41] as shown in ***Eq. 2*** was used to obtain equilibrium properties at zero external pressure ($P$ = 0 GPa), including the static energy $E_0$, volume ($V_0$), and bulk modulus ($B_0$) and its pressure derivate ($B$').

$$E(V) = a + bV^{-2/3} + cV^{-4/3} + dV^{-2} \qquad \textbf{\textit{Eq. 2}}$$

where *a, b, c*, and *d* are fitting parameters.

$F_{el}(V,T)$ is calculated by the following equation [42],

$$F_{el}(V,T) = E_{el}(V,T) - T \cdot S_{el}(V,T) \qquad \textbf{\textit{Eq. 3}}$$

where $E_{el}$ and $S_{el}$ are the internal energy and entropy due to thermal electron excitations, respectively, which can be obtained by the electronic density of states (DOS). Note that the thermal electronic contribution to Helmholtz energy is negligible for non-metals due to the Fermi level in the band gap. $F_{vib}(V,T)$ can be obtained by the following equation [42,43],

$$F_{vib}(V,T) = k_B T \sum_q \sum_j \ln\left\{2\sinh\left[\frac{\hbar\omega_j(q,V)}{2k_B T}\right]\right\} \qquad \textbf{\textit{Eq. 4}}$$

where $\omega_j(q,V)$ represents the frequency of the $j^{\text{th}}$ phonon mode at wave vector $q$ and volume $V$, and $\hbar$ the reduced Planck constant.

### *3.1.2 Details of first-principles calculations*

All DFT-based first-principles and phonon calculations in the present work were performed by the Vienna *ab initio* Simulation Package (VASP) [44]. The projector augmented-wave method (PAW) was used to account for electron-ion interactions in order to increase computational efficiency compared with the full potential methods [45,46]. Electron exchange and correlation effects were described using the generalized gradient approximation (GGA) as implemented by Perdew, Burke, and Ernzerhof (PBE) [47]. The plane-wave basis cutoff energy was 262 eV for structural relaxations and 520 eV for the final static calculations of total energy. The convergent criterion of electronic self-consistency was set as $5\times10^{-6}$ eV/atom for relaxations and static calculations. Seifert et al. [26] reported that $K_3La_5Cl_{18}$ possesses the symmetry of $P6_3/m$ with three Wyckoff sites of 2b, 2c, and 6h. However, the occupancy of the 2b site is less than 1, while the 2c site is occupied by both K and La atoms. Considering these, ATAT [48] was used to search for all possible configurations under these conditions and 9 symmetry inequivalent configurations were found in terms of a 26-atom unit cell. The configuration with the lowest energy was predicted using DFT calculations. Phonon calculations were performed using the supercell method. Table 1 provides detailed settings for DFT-based first-principles and phonon calculations, including reciprocal k-points meshes and supercell sizes for phonon calculations, which ensures the convergence and accuracy of the DFT calculations.

## 3.2 CALPHAD modeling

### *3.2.1 Compounds*

In the present work, the ternary compounds in the LiCl-KCl-$LaCl_3$ system are considered as stoichiometric compounds, including $K_2LaCl_5$ and $K_3La_5Cl_{18}$ (as listed in Sec. 2). Thermodynamic functions of the binary endmembers KCl and $LaCl_3$, are sourced from the JANAF tables [49] and the SSUB database [50]. The Gibbs energy for a given compound is expressed as:

$$G_m = \Delta_f H_m^0(298.15) - T\,S_m^0(298.15) + \int_{298.15}^{T} C_{P,m} dT - T \int_{298.15}^{T} \frac{C_{P,m}}{T} dT \qquad \textbf{\textit{Eq. 5}}$$

where $\Delta_f H_m^0(298.15)$ represents the standard formation enthalpy, $S_m^0(298.15)$ the standard entropy at 298.15 K, and $C_{P,m}$ the heat capacity. For ternary compounds, their thermodynamic data including enthalpy, entropy, and heat capacity are obtained through DFT-based first-principles and phonon calculations, as presented in Sec.4.1.

*3.2.2 Four thermodynamic models for liquid in KCl-LaCl₃*

Regarding the liquid phase, we consider four frequently used models in the literature to describe complex molten salts, including the associate model [7], the two-sublattice ionic model [8], and the MQMQA [9,10] with two sets of coordination numbers.

The species KCl and $LaCl_3$ are assumed in the associate model [7], since no observation of other complex associates exists in the literature. The Gibbs energy of liquid can be expressed as:

$$G_m = y_{KCl}\,{}^{o}G_{KCl}^{Liquid} + y_{LaCl_3}\,{}^{o}G_{LaCl_3}^{Liquid} + RT\left(y_{KCl} ln y_{KCl} + y_{LaCl_3} ln y_{LaCl_3}\right) \qquad \textbf{\textit{Eq. 6}}$$

$$+ y_{KCl} y_{LaCl_3} \sum_{v=0} L_{KCl,LaCl_3}^{v} (y_{KCl} - y_{LaCl_3})^{v}$$

where $y_i$ is the mole fraction of specie $i$ (= KCl or $LaCl_3$), ${}^{o}G_i^{Liquid}$ the Gibbs energy of species $i$, $R$ the gas constant, and $L_{KCl,LaCl_3}^{v}$ the $v^{th}$ interaction parameter between KCl and $LaCl_3$.

In the two-sublattice ionic model [8], the liquid phase in KCl-LaCl$_3$ can be described as:

$$(K^+, La^{3+})_P(Cl^-)_Q \qquad \textbf{\textit{Eq. 7}}$$

where the cations and anions are separated into two sublattices. The site ratios of P and Q follow the following relationships to maintain charge neutrality:

$$P = y_{Cl^-} = 1 \qquad \textbf{\textit{Eq. 8}}$$

$$Q = y_{K^+} + y_{La^{3+}} \qquad \textbf{\textit{Eq. 9}}$$

where $y_i$ represents site fraction of ion $i$, i.e., the mole fraction in each sublattice. The Gibbs energy function can be expressed as:

$$G_m = y_{K^+}{}^oG_{KCl}^{Liquid} + y_{La^{3+}}{}^oG_{LaCl_3}^{Liquid} + RT(y_{K^+}lny_{K^+} + y_{La^{3+}}lny_{La^{3+}}) + {}^{xs}G_m \qquad \textbf{\textit{Eq. 10}}$$

where ${}^{xs}G_m$ represents the excess Gibbs energy and can be described based on the Redlich-Kister polynomial [51], similarly as in ***Eq. 6***:

$${}^{xs}G_m = y_{K^+}y_{La^{3+}} \sum_{v=0} L_{K^+,La^{3+}:Cl^-}^{v}(y_{K^+} - y_{La^{3+}})^v \qquad \textbf{\textit{Eq. 11}}$$

where $L_{K^+,La^{3+}:Cl^-}^{v}$ is the $v^{\text{th}}$ interaction parameter between $K^+$ and $La^{3+}$.

The MQMQA [9,10] describes the KCl-$LaCl_3$ liquid phase by assuming interactions between the quadruplets of $\mathrm{K_2/Cl_2}$, $\mathrm{La_2/Cl_2}$, and $\mathrm{KLa/Cl_2}$. Coordination numbers Z are defined to describe the second nearest neighbors (SNN) coordination number of the species $i$ (= K, La, or Cl) in the quadruplets. Z of anions can be calculated from ***Eq. 12*** to maintain charge neutrality,

$$\frac{q_{\mathrm{K}}}{Z^{\mathrm{La}}_{\mathrm{KLa/ClCl}}} + \frac{q_{\mathrm{La}}}{Z^{\mathrm{La}}_{\mathrm{KLa/ClCl}}} = 2 \times \frac{q_{\mathrm{Cl}}}{Z^{\mathrm{Cl}}_{\mathrm{AB/ClCl}}} \qquad \textbf{\textit{Eq. 12}}$$

where $q_i$ represents the charges of ion $i$ (= K, La, or Cl). Two sets of coordination numbers were applied to KLa/ClCl in the present work as summarized in Table 2. The selection of MQMQA-M3 is based on Sun et al.'s modeling for KCl-$NdCl_3$ [52], while the MQMQA-M4 is based on the MSTDB-TC for KCl-$LaCl_3$ [15].

In MQMQA, the excess Gibbs energy $G^{excess}$ is related to the formation Gibbs energy of the quadruplet, $\Delta g^{ex}_{quadruplet}$,

$$(\mathrm{K_2/Cl_2})_{\mathrm{quad}} + (\mathrm{La_2/Cl_2})_{\mathrm{quad}} = 2(\mathrm{KLa/Cl_2})_{\mathrm{quad}} \qquad \Delta g^{ex}_{\mathrm{AB/Cl_2}} \qquad \textbf{\textit{Eq. 13}}$$

where $\Delta g^{ex}_{\mathrm{KLa/Cl_2}}$ represents the Gibbs energy change when forming the quadruplets and can be described by:

$$\Delta g^{ex}_{\mathrm{KLa/Cl_2}} = \Delta g^{o}_{\mathrm{KLa/Cl_2}} + \sum_{(i+j)\geq 1} g^{ij}_{\mathrm{KLa/Cl_2}} \chi^{i}_{\mathrm{KLa/Cl_2}} \chi^{j}_{\mathrm{LaK/Cl_2}} \qquad \textbf{\textit{Eq. 14}}$$

where $g^{ij}_{\mathrm{KLa/Cl_2}}$ is a function which is dependent of temperature. $\chi^{i}_{\mathrm{KLa/Cl_2}}$ and $\chi^{j}_{\mathrm{LaK/Cl_2}}$ are composition-dependent terms, defined as:

$$\chi^{i}_{\mathrm{KLa/Cl_2}} = \frac{X_{\mathrm{K_2/Cl_2}}}{X_{\mathrm{K_2/Cl_2}} + X_{\mathrm{KLa/Cl_2}} + X_{\mathrm{La_2/Cl_2}}} \qquad \textbf{\textit{Eq. 15}}$$

where $X_{\mathrm{KLa/Cl_2}}$ is the mole fraction of the formula $(\mathrm{KLa/Cl_2})_{\mathrm{quad}}$.

All model parameters in the present work were simultaneously optimized through the Bayesian approach using the Markov Chain Monte Carlo (MCMC) method [18] as implemented in ESPEI [18]. The input data included primarily experimental phase equilibrium data for two or more co-existing phases, mixing enthalpy data, and activity coefficient data from the literature. For stochiometric compounds, their thermochemical data from DFT-based calculations were also used as input. In the present work, each model parameter employed two Markov chains with a standard derivation of 0.1 when initializing its Gaussian distribution. During the modeling process, the chain values can be tracked and the MCMC processes were performed until the model parameters converged.

## 3.3 Bayesian statistics and model selection

ESPEI [18] uses Bayesian parameter estimation to optimize model parameters [53],

$$p(\theta|D,M) = \frac{p(D|\theta,M)p(\theta|M)}{p(D|M)} \qquad \textbf{\textit{Eq. 16}}$$

where $\theta$'s are the model parameters, $M$ the model, and $D$ the input experimental data. In ***Eq. 16***, the posterior $p(\theta|D,M)$ is the probability of model parameters conditioned on data, the likelihood $p(D|\theta,M)$ is the probability that the data are described by parameters, the prior $p(\theta|M)$ contains the domain knowledge in the probability distribution of each parameter, and the marginal likelihood (or evidence) $p(D|M)$ is the probability of data being generated by the model. Here, in CALPHAD modeling, model parameters $\theta$ represent coefficients within Gibbs energy functions. These include terms like $L^{v}_{KCl,LaCl_3}$ from ***Eq. 6***, $L^{v}_{K^+,La^{3+}:Cl^-}$ from ***Eq. 11***, or $g^{ij}_{\mathrm{KLa/Cl_2}}$ from ***Eq. 14***. The model $M$ a consists of Gibbs energy expressions that vary depending on the thermodynamic model used, such as associate model (***Eq. 6***), ionic model (***Eq. 10***), or MQMQA (***Eq. 14***). The input data $D$ includes all properties and datasets provided for fitting the Gibbs energy functions.

Bayesian statistics employed in parameter optimization provide a strategy for model selection for CALPHAD modeling [20,22]. Bayes factor usually suggests which model is more favored by the data, which can be evaluated from the ratio of marginal likelihoods for two competing models:

$$K = \frac{p(D|M_1)}{p(D|M_2)} \qquad \textbf{\textit{Eq. 17}}$$

The marginal likelihood has the desirable qualities of rewarding models that match the data well and penalizing models that are overly complex (i.e., too many degrees of freedom or parameters). The marginal likelihood is determined by:

$$p(D|M) = \int_{\Omega_\theta} p(D|\theta, M)p(\theta|M)d\theta \qquad \textbf{\textit{Eq. 18}}$$

where $\Omega_\theta$ represents the complete parameter space. The evaluation of marginal likelihood requires computation of an integral with dimension given by the number of parameters, which is typically high-dimensional. The evaluation of the marginal likelihood $p(D|M)$ is usually difficult and computationally expensive. The harmonic mean estimator was hence proposed by Newton and Raftery [54] to estimate the marginal likelihood:

$$p(D|M) \approx [\frac{1}{N}\sum_{i=1}^{N} p(D|\theta_i, M)^{-1}]^{-1} \qquad \textbf{\textit{Eq. 19}}$$

where $\theta_i$ are samples from the parameters' prior $p(\theta|M)$. Likelihood values can be obtained from the ESPEI MCMC output, which provides statistical comparison of liquid models through Bayes factor.

# 4 Results and discussion

## 4.1 Thermodynamic properties in LiCl-KCl-$LaCl_3$ by first-principles calculations

Thermodynamic properties of compounds in the KCl-$LaCl_3$ system were predicted using first-principles calculations. Table 3 summarizes the equilibrium properties of $V_0$, $B_0$, and B' at 0 K obtained by DFT-based calculations in comparison with experiments. The present work predicts the bulk modulus $B_0$ value to be 16.23 GPa for KCl, which is slightly lower than the experimental measurement of 19.7 GPa by Norwood et al. [55]. The equilibrium volume $V_0$ of $LaCl_3$ is predicted to be 27.37 Å$^3$/atom in the present work, which is in good agreement with the measured 26.38 Å$^3$/atom by Zachariasen [56]. It indicates that the present DFT calculations provide reliable predictions regarding the equilibrium properties of compounds in the KCl-$LaCl_3$ system. The present DFT calculations predicted the $V_0$ of $K_2LaCl_5$ to be 29.96 Å$^3$/atom and $B_0$ to be15.89 GPa. For $K_3La_5Cl_{18}$, the $V_0$ is reported to be 27.49 Å$^3$/atom and $B_0$ to be 26.46 GPa, respectively.

Thermodynamic properties at finite temperature are obtained through DFT-based QHA in terms of phonon. Figure 1 compares the predicted values of heat capacity ($C_p$), entropy (S), and enthalpy (H-$H_{300}$) of KCl and $LaCl_3$ to those from the SGTE database [50]. For KCl, the present QHA results are slightly higher than the SGTE data [50] with a minor difference of about 6% for S. For $LaCl_3$, the present QHA results slightly underpredict Cp and H-$H_{300}$ compared to SGTE [50], particularly at higher temperature. The differences in Cp for $LaCl_3$ remain less than 3.2 J/mol-atom-K at high temperature, while the entropy and enthalpy closely match the SGTE values [50]. Figure 2 shows the predicted $C_p$ of $K_2LaCl_5$ and $K_3La_5Cl_{18}$ in comparison with experiments [27,28], demonstrating an excellent agreement. For example, at 500 K, $C_p$ of $K_2LaCl$ from QHA calculations is 26.99 J/mol-K, which is 0.8% and 0.5% higher than 26.77 J/mol-K reported by Reuter et al. [27] and 0. 26.86 J/mol-K by Gaune-Escard et al. [28], respectively. For $K_3La_5Cl_{18}$, the predicted $C_p$ at 500 K is 25.99 J/mol-K, slightly lower than 26.34 J/mol-K by Reuter et al. [27]. These thermodynamic data of $K_2LaCl_5$ and $K_3La_5Cl_{18}$ obtained from the present DFT calculations are used in the present CALPHAD modeling.

### 4.2 Model selection for the liquid phase in the KCl-$LaCl_3$ system

The KCl-$LaC_3$ system is modeled using four models, i.e., the associate model (Associate-M1), the ionic model (Ionic-M2), and the MQMQA-M3 and the MQMQA-M4 with different coordination numbers. Table 4 summarizes the modeling parameters for these models. Each model has four adjustable parameters, which were optimized by at least 1000 MCMC iterations. This process continued until the posterior probability values from each Markov chain stabilized, i.e., the model parameters converged. Figure 3 compares the phase diagrams calculated from these four models with experimental data [26,31]. It shows that for the liquidus of the KCl-rich region, MQMQA-M3 and MQMQA-M4 provide better agreements with experimental data than Associate-M1 and Ionic-M2. Table 5 lists the invariant reactions predicted by different models in comparison with experimental data [26,31]. In general, all these four models show excellent agreements with the experimental data. Ionic-M2 and MQMQA-M4 slightly overpredict the invariant temperatures compared to those from Asscociate-M1 and MQMQA-M3. Specifically, Ionic-M2 and MQMQA-M4 predict a eutectic temperature of 854 K for the reaction of Liquid↔KCl+$K_2LaCl_5$, which is 9 K higher than 845 K reported by Song et al. [31] and 1 K above 853 K by Seifert et al. [26]. For the melting temperature of $K_2LaCl_5$, Ionic-M2 predicts a 917 K, while MQMQA-M4 predicts a 926 K, both are higher than the measured 913 K by Seifert et al.[26] and 916 K by Song et al. [31]. Associate-M1 slightly underpredicts the peritectic temperature of the reaction Liquid+$LaCl_3$↔$K_3La_5Cl_{18}$ at 882 K, which is 3 K lower than the 885 K by Seifert et al. [26]. The mean absolute error (MAE) for predicting these invariant temperatures using Associate-M1 is 2.5 K. MQMQA-M3 provides a good agreement with experimental data, with a slightly lower prediction of eutectic temperature for the reaction Liquid↔$K_2LaCl_5$+$K_3La_5Cl_{18}$ at 845 K, which is 6 K lower than 851 K reported by Seifert et al. [26]. Regarding the invariant compositions x($LaCl_3$), MQMQA-M3 and MQMQA-M4 offer better predictions with an MAE of 0.017 for both, compared to 0.025 for Associate-M1 and 0.022 for Ionic-M2.

Figure 4 shows the values of mixing enthalpy of liquid at 1173 K calculated using these four models, which are compared with experimental data by Papatheodorou and Ostvold [29]. The comparison indicates that Associate-M1 and Ionic-M2 predict lower mixing enthalpy values than those by MQMQA-M3 and MQMQA-M4. For example, at x($LaCl_3$) = 0.496, Associate-M1 predicts -15.469 kJ/mol and Ionic-M2 predicts -15.455 kJ/mol, slightly lower than the -15.097 kJ/mol predicted by MQMQA-M3 and -15.118 kJ/mol by MQMQA-M4. When compared to the mixing enthalpy value of -15.319 kJ/mol reported by Papatheodorou and Ostvold [29], Associate-M1 and Ionic-M2 show a closer alignment, with a difference of 0.9%, compared to a 1.4% difference for MQMQA-M3 and MQMQA-M4. Notably, Associate-M1 and Ionic-M2 align more closely with experimental data [29], particularly in the composition region with x($LaCl_3$) > 0.4.

According to phase diagrams (Figure 3) and mixing enthalpy predictions (Figure 4), each model demonstrates a strength in predicting thermodynamic properties of the KCl-$LaCl_3$ system. Associate-M1 and Ionic-M2 perform well in predicting mixing enthalpy, showing a better match with experimental data as discussed above. However, these two models are less accurate in predicting phase boundaries compared to MQMQA-M3 and MQMQA-M4, especially at the $LaCl_3$-rich region. Choosing an appropriate model remains challenging, as it requires balancing agreement with all available experimental data. A quantitative method is hence needed to determine the overall favorability for the models of study.

Bayesian parameter estimation through MCMC offers a powerful tool to statistically compare models, as discussed in Sec.3.3. Table 6 lists the estimated marginal likelihood value for each model, indicating that MQMQA-M3 has the highest marginal likelihood value of $ln(p(D|M_3))$ = -370.628. Consequently, the Bayes factors for the other models were calculated with respect to MQMQA-M3, using the marginal likelihood value of MQMQA-M3 as the numerator shown in ***Eq. 17***. The interpretation of the Bayes factor is also included in Table 6 according to the guideline by Kass and Raftery [21]. That is, a $log_{10}$(Bayes factor) value between 0 and 1/2 suggests not worth more than a bare mention in comparing two models even if it points very slightly towards the model at the denominator, a range of 1/2 to 1 indicates substantial evidence in favor of the model at the denominator, 1 to 2 denotes a strong evidence, and a value greater than 2 represents a decisive evidence in favor of the model at the denominator.

It can be seen in included in Table 6 that the Bayes factor $log_{10}K_{M3/M1}$ is 28.016, suggesting decisive evidence of favoring MQMQA-M3 over Associate-M1. This is due to the larger discrepancy in phase boundary predictions from Associate-M1 compared to the experimental phase diagram data shown in Figure 3(a). Regarding the Ionic-M2, the Bayes factor $log_{10}K_{M3/M2}$ = 1.208 indicates strong evidence of favoring MQMQA-M3 over Ionic-M2. Additionally, the Bayes factor comparing the two MQMQA models, $log_{10}K_{M3/M4}$, is 0.344, suggesting that MQMQA-M3 is slightly favored over MQMQA-M4, but not worth a bare mention when comparing these two models. It implies that the choice of these two sets of coordination numbers did not significantly affect the performance of MQMQA models in predicting thermodynamic properties in the KCl-$LaCl_3$ system.

In summary, Bayesian parameter estimation through MCMC indicates that MQMQA-M3 is more favored, in terms of the input data of phase boundary and mixing enthalpy, over the other three models. This approach provides a robust technique for estimating the marginal likelihood values to assess the probability of data being generated by the model and calculating Bayes factors to statistically compare different models.

## 4.3 Thermodynamic modeling of the LiCl-KCl-$LaC_3$ system

The MQMQA-M3 is selected for the KCl-$LaCl_3$ system according to the Bayes factor. The ternary LiCl-KCl-$LaCl_3$ system is further improved by this model. The other two binary systems of LiCl-KCl and LiCl-$LaCl_3$ are taken from the MSTDB-TC [15]. The available experimental data used for CALPHAD modeling include phase boundary data measured by Nakamura et al. [34] and Venkata Krishnan et al. [35], activity coefficient data measured by Bagri and Simpson [32] and Samin et al. [33], and mixing enthalpy data of $LaCl_3$ in LiCl-KCl eutectic [36]. Table 7 summarizes the modeled ternary interaction parameters after MCMC optimizations.

Thermodynamic properties predicted from the present CALPHAD modeling are compared with available experimental data. In addition, uncertainty quantification (UQ) is performed to propagate parameter uncertainties into property predictions. Figure 5(a) presents the values of activity coefficient of $LaCl_3$ in the eutectic LiCl-KCl at 773 K, compared with the measurements by Bagri et al. [32] and Samin et al. [33]. The present modeling aligns more closely with the results by Bagri et al. [32], since they provided a larger dataset over a broader composition range than those by Samin et al. [33]. The present CALPHAD model shows a good agreement in the low x(LaCl3) region (x(LaCl3) < 0.015) but slightly overestimates the activity coefficients when x(LaCl3) increases to 0.02. The MAE of the present modeling, compared with values reported by Bagri et al. [32], is 0.0016. The present modeling represents a significant improvement over the previous work by Hao et al. [14], which had a MAE of 0.0078 for the activity coefficients. While MSTDB-TC [15] shows a good agreement in the composition range of x(LaCl3) from 0.022 to 0.027, it has an overall MAE of 0.0024 compared to the values reported by Bagri et al. [32]. The present UQ values were performed using the last 10 MCMC iterations with 60 MCMC samples. The shadow region in Figure 5(b) illustrates the uncertainties in predicting activity coefficients, with a 95% credible interval (Bayesian credible intervals containing 95% of the activity coefficients samples). This indicates that the present model has an uncertainty range of -10% to +2% in predicting the activity coefficient values.

Figure 6 shows the mixing enthalpy at 873 K and 1133 K calculated from the present modeling, in comparison with experimental data [36] and modeling results by Hao et al. [14] and MSTDB-TC [15]. In Figure 6(a), at 873 K, all these three modeling works closely match experimental measurements [36] for x($LaCl_3$) < 0.2. For example, at x($LaCl_3$) = 0.1926, the present modeling predicts a mixing enthalpy of -4231 J/mol, which is 51 J/mol lower than the experimental value [36] of -4180 J/mol. In comparison, Hao et al. [14] predicts -4266 J/mol and MSTDB-TC [15] predicts -4659 J/mol, showing larger differences of 86 J/mol and 389 J/mol, respectively. The MAE of the present modeling in predicting mixing enthalpy at 873 K is 101 J/mol, compared to 120 J/mol for Hao et al. [14] and 312 J/mol for MSTDB-TC [15], highlighting the improved accuracy of the present modeling. Figure 6(a) also shows different shapes of the curves in the $LaCl_3$-rich region predicted by three modeling works. The present modeling predicts a minimum energy similar to that by MSTDB-TC [15] at around -5650 J/mol, whereas Hao et al. [14] predicts a value of -6186 J/mol, which is around 540 J/mol lower. Note that experiments investigations at 1133 K primarily focus on the $LaCl_3$-rich region [36].

Figure 6(b) indicates that both Hao et al. [14] and MSTDB-TC [15] predict lower values of mixing enthalpy compared to the present modeling and experiments [36]. The present work improves the accuracy of mixing enthalpy predictions, reducing the MAE to 45.84 J/mol, compared to 159.75 J/mol by Hao et al. [14] and 183.30 J/mol by MSTDB-TC [15]. Figure 6(c) presents the uncertainty quantification of the present modeling in predicting mixing enthalpy, represented by the 95% credible interval. At 873 K, the uncertainty range in predicting mixing enthalpy is around -5% to +5%. At 1133 K, the uncertainty range in predicting mixing enthalpy is around -6% to +6% with all existing experimental data [36] falling within the lower boundary of the uncertainty region. This implies that the present modeling might underestimate the mixing enthalpy, particularly around x(LaCl3) = 0.4. Further experiments or simulations in this composition range are recommended to enhance the accuracy of the present CALPHAD modeling. Figure 7 shows the predicted fraction of each quadruplet in the liquid phase. The peak fractions of the LiLa/ClCl and LaK/ClCl quadruplets appear around x($LaCl_3$) = 0.4, indicating strong short-range ordering (SRO) and consistent with the lowest mixing enthalpy around x($LaCl_3$) = 0.4.

Figure 8 shows the partial isopleth between the eutectic KCl-LiCl and $LaCl_3$ calculated from the present CALPHAD modeling compared with experimental measurements [34,35], depicting a close match of liquidus and solidus lines. For the eutectic reaction Liquid↔KCl+LiCl+$K_2LaCl_5$, the present modeling predicts a eutectic temperature of 630 K, which is 5 K higher than the 625 K reported by Nakamura et al. [34] and Venkata Krishnan et al. [35]. Similarly, for the reaction Liquid↔LiCl+$K_2LaCl_5$+ $K_3La_5Cl_{18}$, the present modeling predicts the eutectic temperature at 705

K, which is 3 K higher than the reported 702 K by Nakamura et al. [34] and Venkata Krishnan et al. [35]. Overall, the present modeling of LiCl-KCl-$LaCl_3$ demonstrates good agreement with experimental data [32–35] regarding phase boundary properties.

## 5 Conclusions

The present work demonstrates an application of Bayesian model selection to identify optimal model for molten salts, focusing on the KCl-$LaCl_3$ system. Four candidate models are considered, which are the associate model, the two-sublattice ionic model, and two MQMQA models with different coordination numbers. By estimating the marginal likelihoods of each model from MCMC optimization and calculating Bayes factors, one of the MQMQA models is suggested as the most favorable one to describe the KCl-$LaCl_3$ system based on available input data. Additionally, DFT-based calculations provide important thermodynamic properties for compounds in KCl-$LaCl_3$, including equilibrium volumes, bulk moduli, enthalpies, entropies, and heat capacities. Furthermore, the ternary LiCl-KCl-$LaCl_3$ system is modelled, demonstrating a better agreement with experimental data compared to the previous CALPHAD modeling works in the literature. The uncertainty quantification and propagation show that the present modeling provides reliable predictions of activity coefficients and mixing enthalpy when compared with experimental data. The present work indicates that the Bayesian model selection approach facilitates a more rational comparison among different liquid models, enhancing the accuracy of thermodynamic predictions in molten salts.

## Acknowledgments

The authors acknowledge financial supports by the U.S. Department of Energy, Office of Nuclear Energy's Nuclear Energy University Programs via Award No. DE-NE0009288. CC, KT and PAB acknowledge with thanks funding from EPSRC (EP/X011607/1). First-principles calculations were performed partially on the Roar supercomputer at the Pennsylvania State University's Institute for Computational and Data Sciences (ICDS), partially on the resources of the National Energy Research Scientific Computing Center (NERSC) supported by the U.S. Department of Energy, Office of Science User Facility operated under Contract No. DE-AC02-05CH11231, and partially on the resources of ACCESS (previously the Extreme Science and Engineering Discovery Environment, XSEDE) supported by National Science Foundation (NSF) with Grant No. ACI-1548562.

# 6 Figures and Figure Captions

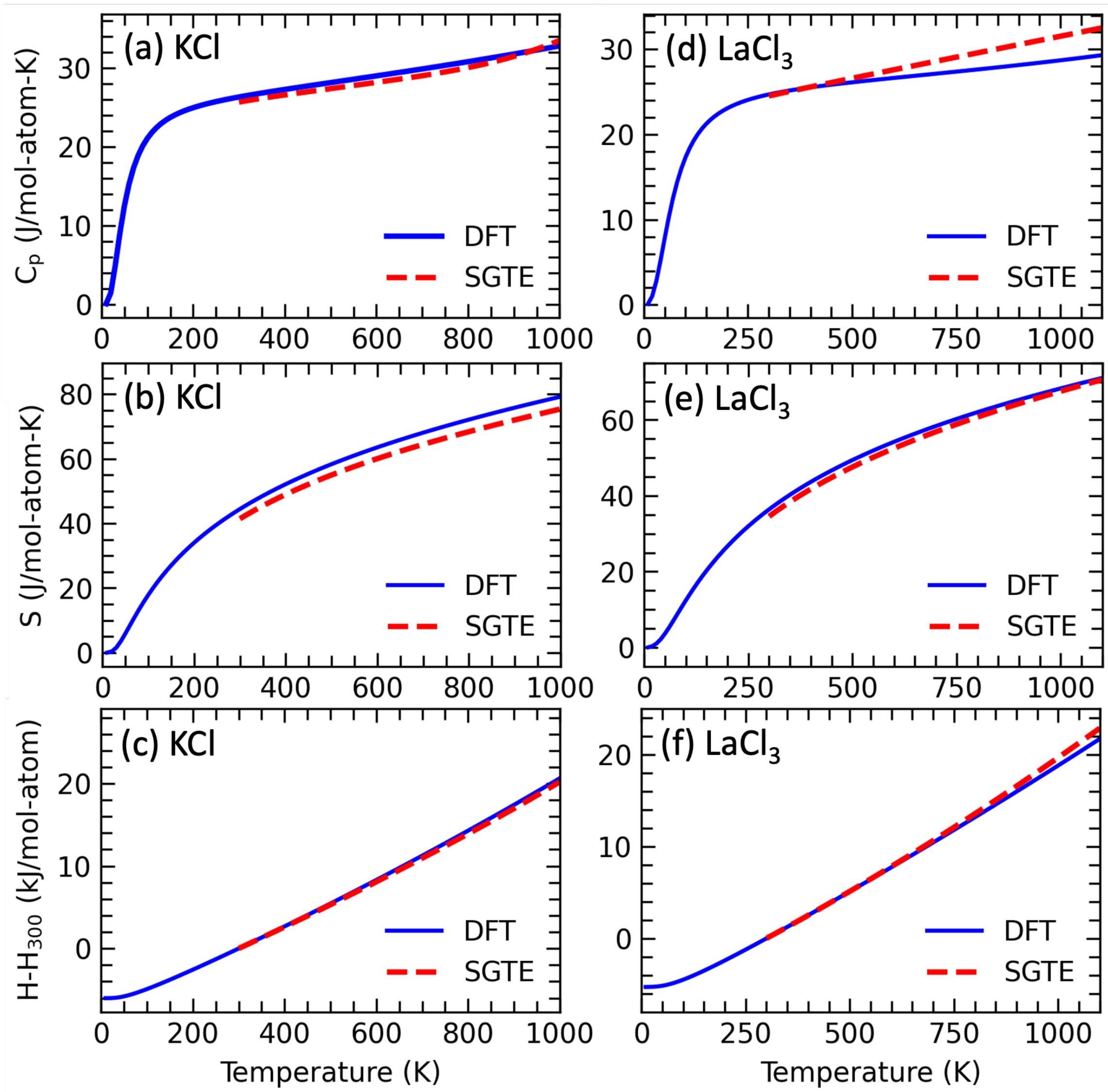

Figure 1. Comparison of the present values (blue lines) of heat capacity $C_p$, entropy S, and enthalpy with reference to the value at 300 K ($H\text{-}H_{300}$) for (a-c) KCl and (d-f) $LaCl_3$ from the DFT-based phonon calculations (blue lines) with the SGTE data [50] (red dash lines).

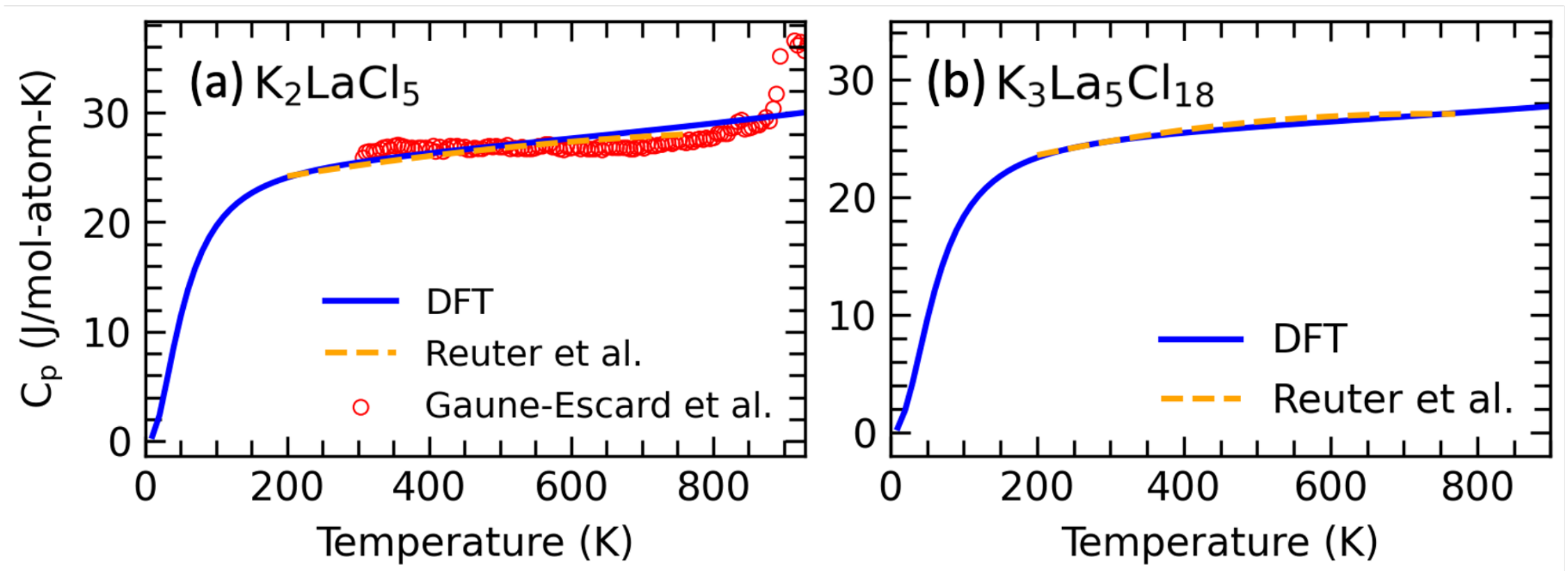

Figure 2. Comparison of heat capacity $C_p$ values for (a) $K_2LaCl_5$ and (b) $K_3La_5Cl_{18}$ from the present DFT-based QHA (blue lines) with experiments by Reuter et al. [27] (yellow dash lines) and Gaune-Escard et al. [28] (red circles).

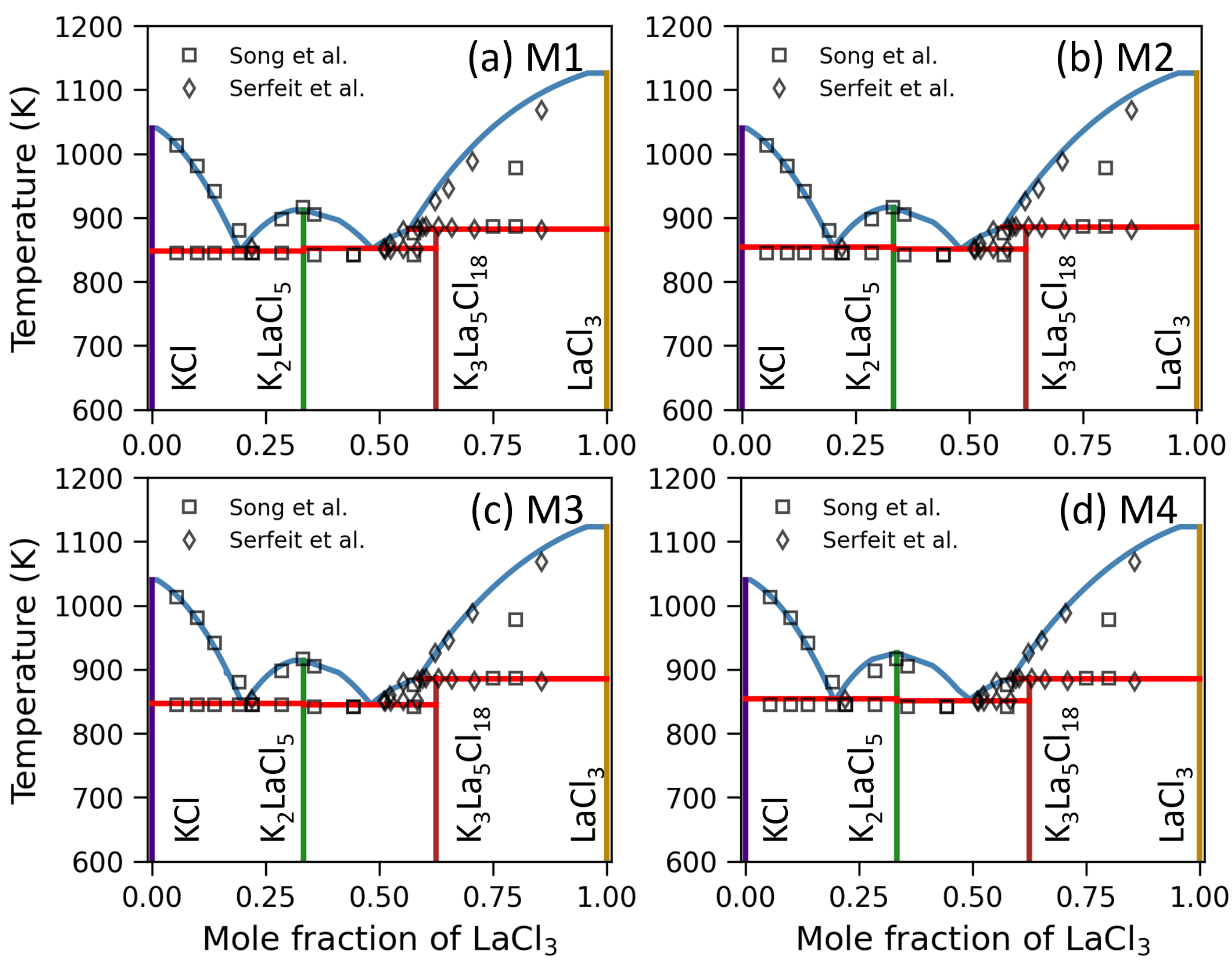

Figure 3. Predicted phase diagrams of the $LaCl_3$-KCl system with the liquid phase modeled using (a) associate model (Associate-M1), (b) ionic model (Ionic-M2), (c) MQMQA (MQMQA-M3), and (d) MQMQA (MQMQA-M4) in comparison to experimental data [26,31].

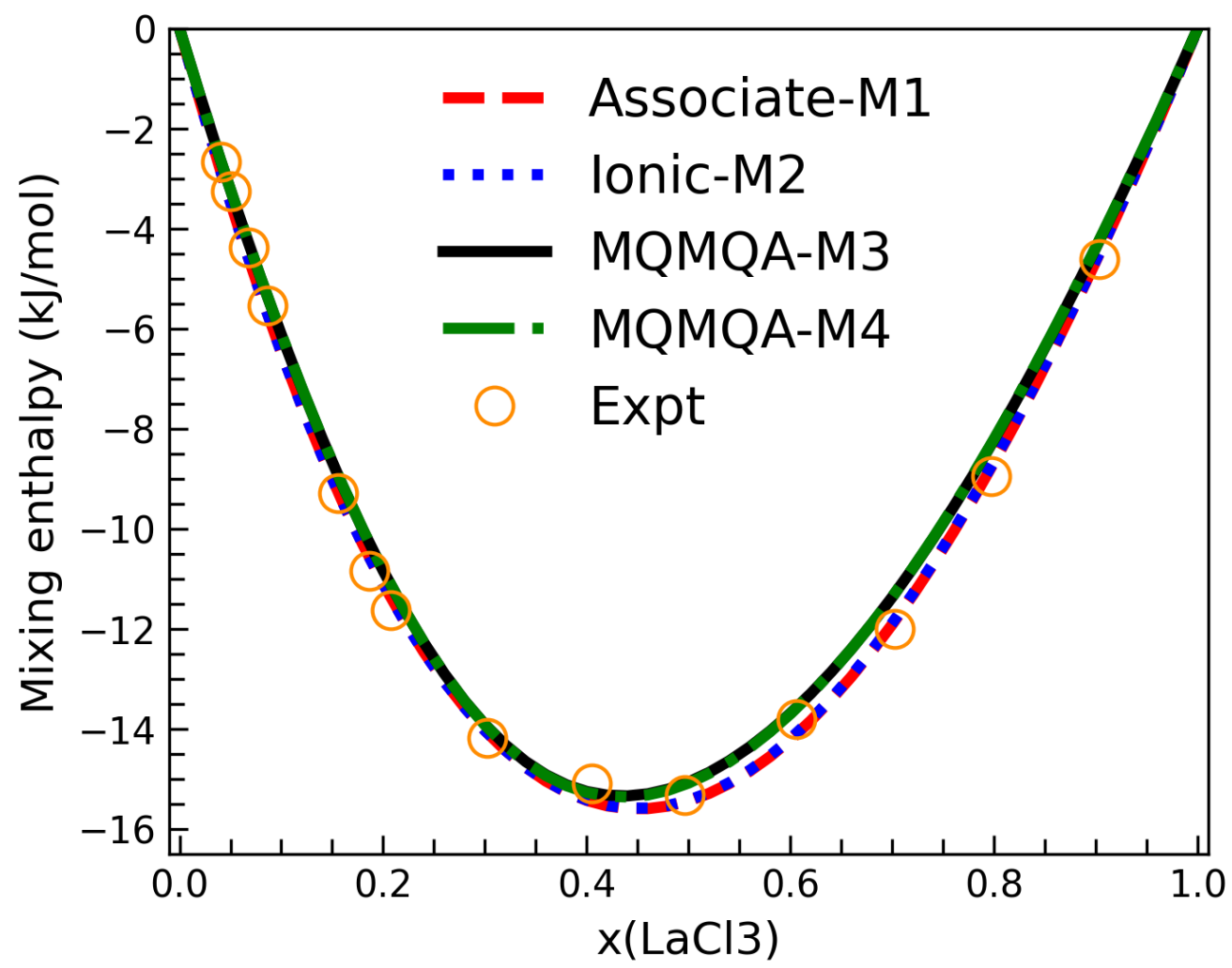

Figure 4. Predicted values of mixing enthalpy of the $LaCl_3$-KCl liquid at 1173K from four models in comparison with experimental data by Papatheodorou and Ostvold [29]. The red dash line represents Associate-M1, the blue dotted line represents Ionic-M2, the black solid line represents MQMQA-M3, and the green dash dotted line represents MQMQA-M4.

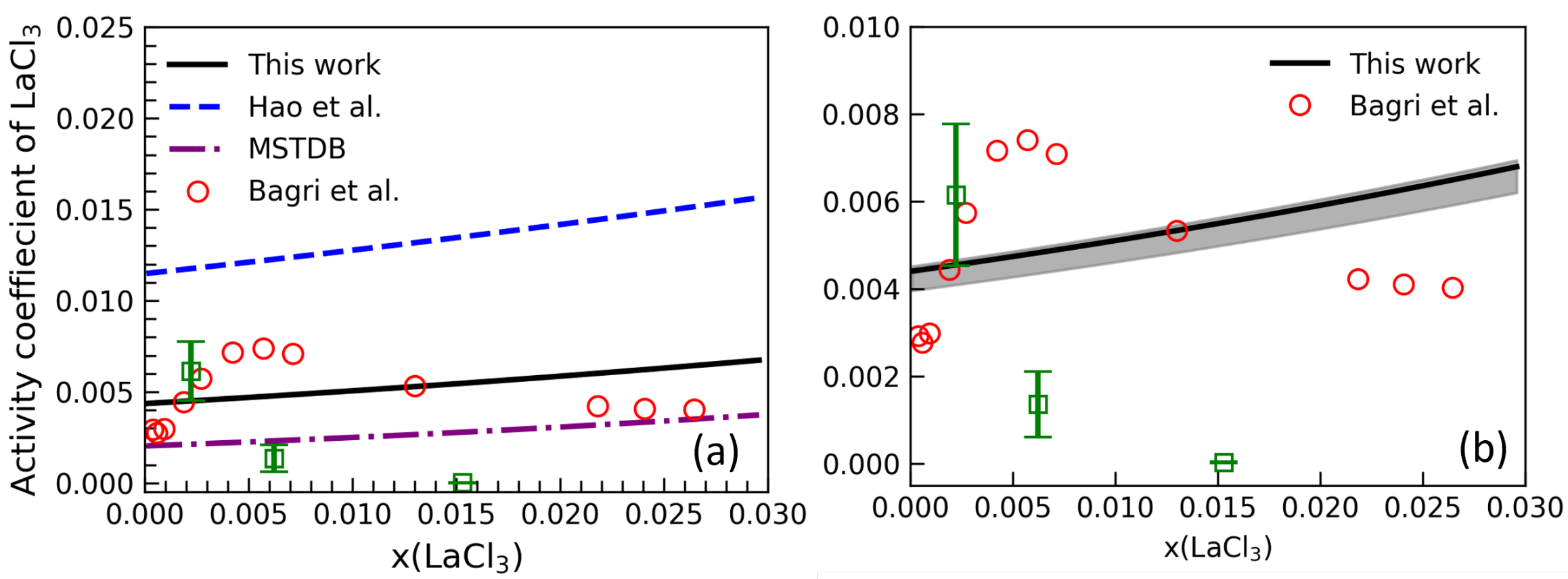

Figure 5. (a) Activity coefficients of $LaCl_3$ in KCl-LiCl eutectic at 773 K calculated from the present modeling in comparison with the modeling works from Hao et al. [14] (blue dash line) and MSTDB-TC [15] (purple dash dotted line), experimental measurements by Bagri et al. (red circles) and Samin et al. (green squares). (b) The uncertainty of the present modeling in predicting activity coefficients shown in the grey region using a 95% credible interval in predicting the values.

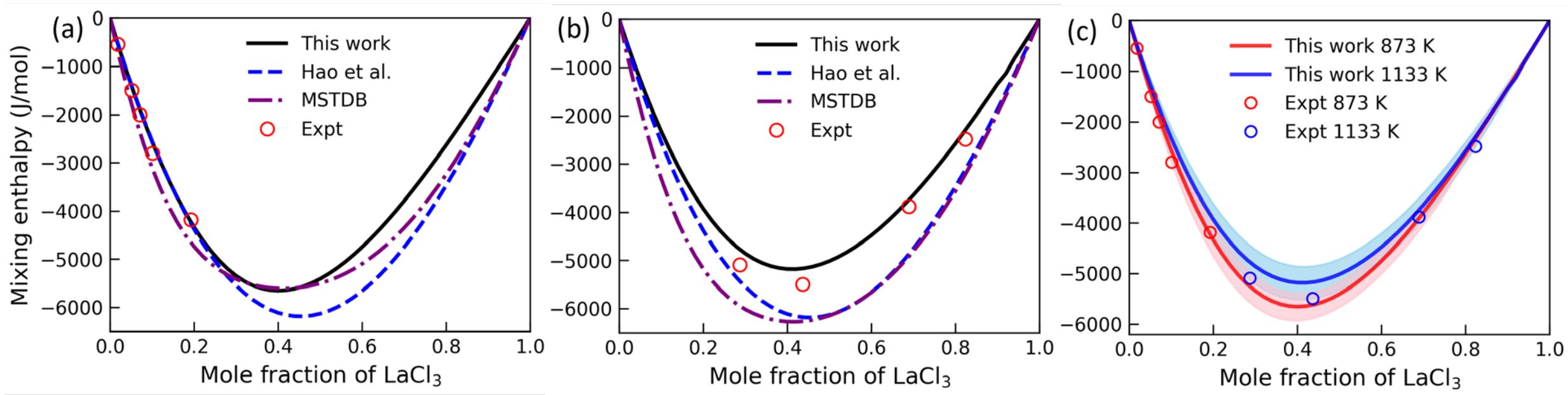

Figure 6. Mixing enthalpy in LiCl-KCl-$LaCl_3$ at (a) 873 K and (b) 1133 K calculated from the present modeling compared with experimental measurements by Goncharov et al. [36] and the previous modeling works by Hao et al. [14] and MSTDB-TC [15]. (c) The uncertainty of the present modeling in predicting mixing enthalpy shown in the light red region (for 873 K) and the light blue region (for 1133 K) using a 95% credible interval.

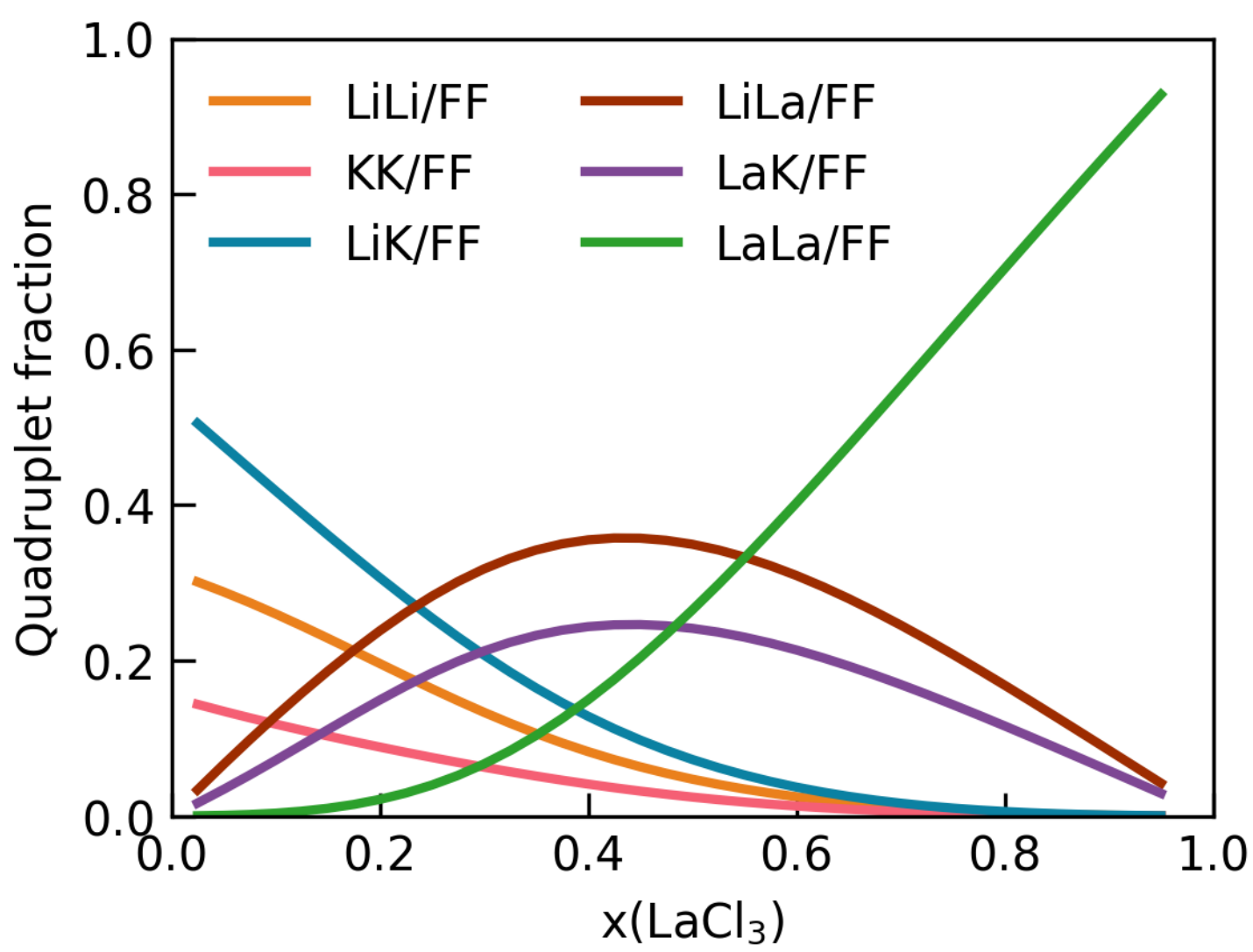

Figure 7. Precited quadruplet fractions in the LiCl-KCl-$LaCl_3$ liquid at 1133 K according to the present CALPHAD modeling using MQMQA.

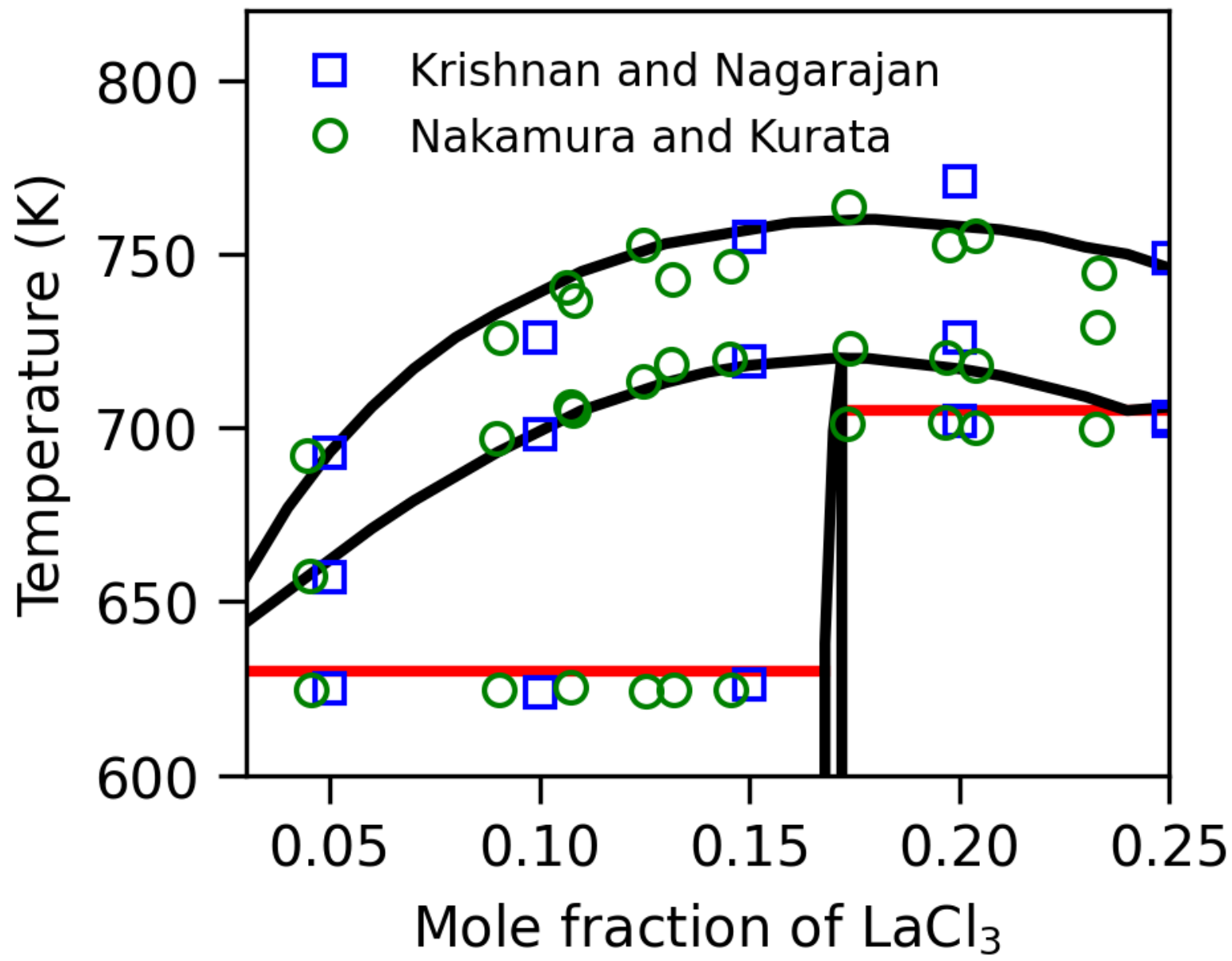

Figure 8. Partial isopleth between the eutectic KCl-LiCl and $LaCl_3$ calculated from the present modeling work in comparison with experimental measurements by Krishnan and Nagarajan (blue squares) [35] and Nakamura and Kurata (green circles) [34].

## 7 Tables and Table Captions

Table 1. Details of the present DFT-based first-principles and phonon calculations for each compound, including space group, total atoms in the supercells, k-point meshes for structure relaxations and the final static calculations (indicated by DFT), supercell sizes for phonon calculations, and k-point meshes for phonon calculations.

| Phase | Space Group | Atoms in crystallographic cell | k-points for DFT | Atoms in supercell for phonon | k-points for phonon |
|---|---|---|---|---|---|
| KCl | $Fm\bar{3}m$ | 8 | 8×8×8 | 64 | 3×3×3 |
| $LaCl_3$ | $P6_3/m$ | 8 | 8×8×12 | 64 | 2×2×2 |
| $K_2LaCl_5$ | Pnma | 32 | 7×7×4 | 32 | 4×4×2 |
| $K_3La_5Cl_{18}$ | P3 | 26 | 8×8×5 | 26 | 5×5×3 |

Table 2. Coordination numbers used in the present CALPHAD modeling with MQMQA for the liquid phase.

| A | B | | $Z^{\mathrm{A}}_{\mathrm{AB/ClCl}}$ | $Z^{\mathrm{B}}_{\mathrm{AB/ClCl}}$ | $Z^{\mathrm{F}}_{\mathrm{AB/ClCl}}$ |
|---|---|---|---|---|---|
| $K^+$ | $K^+$ | | 6.0 | 6.0 | 6.0 |
| $La^{3+}$ | $La^{3+}$ | | 6.0 | 6.0 | 2.0 |
| $K^+$ | $La^{3+}$ | MQMQA-M3 | 2.0 | 6.0 | 2.0 |
| | | MQMQA-M4 | 3.5 | 6.0 | 2.55 |

Table 3. Predicted equilibrium properties of volume $V_0$, bulk modulus $B_0$, and the first derivative of bulk modulus with respect to pressure $B'$ for compounds in the $KCl$-$LaCl_3$ system based on the present EOS fitting at 0 K (***Eq. 2***). Experimental data are also listed for comparison.

| Compounds | $V_0$ (Å$^3$/atom) | $B_0$ (GPa) | $B'$ | References |
|---|---|---|---|---|
| KCl | 32.62 | 16.23 | 4.67 | This work |
| | | 19.7 | | Norwood et al. [55] |
| $LaCl_3$ | 27.37 | 29.02 | 6.40 | This work |
| | 26.38 | | | Zachariasen [56] |
| $K_2LaCl_5$ | 29.96 | 15.89 | 5.38 | This work |
| $K_3La_5Cl_{18}$ | 27.49 | 26.46 | 6.35 | This work |

Table 4. Details of four models and the resulting model parameters in the present work for $KCl$-$LaCl_3$.

| Name | Models | Model parameters |
|---|---|---|
| Associate-M1 | $(KCl, LaCl_3)$ | $L^0_{KCl,LaCl_3} = -61777.370 + 1.218 * \mathrm{T}$ <br> $L^1_{KCl,LaCl_3} = -12634.241 - 1.383 * \mathrm{T}$ |
| Ionic-M2 | $(K^+, La^{3+})_P(Cl^-)_Q$ | $L^0_{K^+,La^{3+}:Cl^-} = -61720.724 + 5.247 * \mathrm{T}$ <br> $L^1_{K^+,La^{3+}:Cl^-} = -12786.004 + 3.360 * \mathrm{T}$ |
| MQMQA-M3 | See Table 2 | $\Delta g^{ex}_{\mathrm{KLa/Cl_2}} = -13295.794 - 1.236 * \mathrm{T} + (-9228.101)\chi_{\mathrm{KLa/Cl_2}} + (-2071.866)\chi_{\mathrm{LaK/Cl_2}}$ |
| MQMQA-M4 | See Table 2 | $\Delta g^{ex}_{\mathrm{KLa/Cl_2}} = -13313.121 - 1.333 * \mathrm{T} + (-9173.701)\chi_{\mathrm{KLa/Cl_2}} + (-1989.647)\chi_{\mathrm{LaK/Cl_2}}$ |

Table 5. Predicted invariant equilibria in the KCl-$LaCl_3$ system by the four models, compared with experimental data.

| Reaction | | $x(LaCl_3)$ | Temperature (K) | Source |
|---|---|---|---|---|
| Eutectic | Liquid↔KCl+$K_2LaCl_5$ | 0.22 | 853 | Seifert et al.[26] |
| | | 0.22 | 845 | Song et al. [31] |
| | | 0.197 | 848 | Associate-M1 |
| | | 0.202 | 854 | Ionic-M2 |
| | | 0.204 | 847 | MQMQA-M3 |
| | | 0.200 | 854 | MQMQA-M4 |
| Melting | Liquid↔$K_2LaCl_5$ | 0.333 | 913 | Seifert et al.[26] |
| | | 0.333 | 916 | Song et al. [31] |
| | | 0.333 | 913 | Associate-M1 |
| | | 0.333 | 917 | Ionic-M2 |
| | | 0.333 | 915 | MQMQA-M3 |
| | | 0.333 | 926 | MQMQA-M4 |
| Eutectic | Liquid↔$K_2LaCl_5$+$K_3La_5Cl_{18}$ | 0.51 | 851 | Seifert et al.[26] |
| | | 0.485 | 852 | Associate-M1 |
| | | 0.481 | 851 | Ionic-M2 |
| | | 0.482 | 845 | MQMQA-M3 |
| | | 0.49 | 851 | MQMQA-M4 |
| Peritectic | Liquid+$LaCl_3$↔$K_3La_5Cl_{18}$ | 0.595 | 885 | Seifert et al.[26] |
| | | 0.565 | 882 | Associate-M1 |
| | | 0.573 | 885 | Ionic-M2 |
| | | 0.586 | 885 | MQMQA-M3 |
| | | 0.586 | 885 | MQMQA-M4 |

Table 6. Summary of the predicted marginal likelihood and Bayes factor for each model with respect to those of MQMQA-M3, and the suggested strength of evidence according to Kass and Raftery [21].

| Model name | Marginal likelihood (ln value) | Bayes factor $\log_{10}K$ | Strength of evidence |
|---|---|---|---|
| Associate-M1 | -435.136 | 28.016 | Decisive |
| Ionic-M2 | -373.410 | 1.208 | Strong |
| MQMQA-M3 | -370.628 | - | - |
| MQMQA-M4 | -371.420 | 0.344 | Not worth more than a bare mention |

Table 7. Modelled ternary interaction parameters in the LiCl-KCl-$LaCl_3$ liquid using MQMQA after MCMC optimization in the present work.

| System | Ternary parameters |
|---|---|
| LiCl-KCl-$LaCl_3$ | $\Delta g^{101}_{\mathrm{LaLi(K)/Cl_2}} = 10153.591$<br>$\Delta g^{001}_{\mathrm{KLI(La)/Cl_2}} = 18921.383 - 12.806 * \mathrm{T}$ |